\documentclass[aps,prb,twocolumn,showpacs,showkeys,floatfix]{revtex4}

\usepackage{graphicx,color}
\graphicspath{{figs/}}
\begin{document}
\title{Anisotropic Young's Modulus for Single-Layer Black Phosphorus: The Third Principle Direction Besides Armchair and Zigzag}
\author{Jin-Wu Jiang}
    \altaffiliation{Corresponding author: jwjiang5918@hotmail.com}
    \affiliation{Shanghai Institute of Applied Mathematics and Mechanics, Shanghai Key Laboratory of Mechanics in Energy Engineering, Shanghai University, Shanghai 200072, People's Republic of China}

%\date{22 December 2009}
\date{\today}
\begin{abstract}

We derive an analytic formula for the Young's modulus in single-layer black phosphorus using the valence force field model. By analyzing the directional dependence for the Young's modulus, we explore the third principle direction with direction angle $\phi_{\rm tp}=0.268\pi$ besides armchair and zigzag directions. The maximum Young's modulus value is in the third principle direction. More specifically, the Young's modulus is 52.2~{Nm$^{-1}$}, 85.4~{Nm$^{-1}$}, and 111.4~{Nm$^{-1}$} in the armchair direction, zigzag direction, and the third principle direction, respectively. This new principle direction is of significance for future discussions of other anisotropic properties in the single-layer black phosphorus.

\end{abstract}

\maketitle
\pagebreak

%\section{Introduction}
Few-layer black phosphorus (BP) is another interesting quasi two-dimensional system that has recently been explored as an alternative electronic material to graphene, boron nitride, and the transition metal dichalcogenides for transistor applications\cite{LiL2014,LiuH2014,BuscemaM2014,BuscemaM2014nc}. This initial excitement surrounding BP is because unlike graphene, BP has a direct bandgap that is layer-dependent.  Furthermore, BP also exhibits a carrier mobility that is larger than MoS$_{2}$\cite{LiuH2014}. The van der Waals effect in bulk BP was discussed by Appalakondaiah et.al.\cite{AppalakondaiahS2012prb} First-principles calculations show that single-layer BP (SLBP) has a band gap around 0.8~{eV}, and the band gap decreases with increasing thickness.\cite{DuY2010jap,LiuH2014} For SLBP, the band gap can be manipulated via mechanical strain in the direction normal to the BP plane, where a semiconductor-metal transition was observed.\cite{RodinAS2014,PengXH2014prb,GuoH2014jpcc}

The single-layer BP has a characteristic puckered structure, which leads to the two anisotropic in-plane directions. As a result of this puckered configuration, anisotropy has been found in various properties for the single-layer BP, such as the optical properties,\cite{XiaF2014nc,TranV2014prb,LowT2014prb} the electrical conductance,\cite{FeiR2014nl} the mechanical properties,\cite{AppalakondaiahS2012prb,QiaoJ2014nc,JiangJW2014bpyoung,QinGarxiv14060261,WeiQ2014apl} and the Poisson's ratio.\cite{JiangJW2014bpnpr,QinGarxiv14060261,JiangJW2014bpsw}

The present work focuses on the Young's modulus of the SLBP. In all existing works, the Young's modulus in the armchair direction is much less than that in the zigzag direction. For instance, the Young's modulus from the {\it ab initio} calculations for the armchair and zigzag-directions is 28.9~{Nm$^{-1}$} and 101.6~{Nm$^{-1}$} in Ref.\onlinecite{QiaoJ2014nc}, or 21.9~{Nm$^{-1}$} and 56.3~{Nm$^{-1}$} in Ref.\onlinecite{JiangJW2014bpyoung}, or 19.5~{Nm$^{-1}$} and 78.0~{Nm$^{-1}$} in Ref.\onlinecite{JiangJW2014bpsw}.

From the above, in most existing works, the investigation of anisotropic properties is usually performed by comparing these properties in two principle directions, i.e., armchair and zigzag directions. However, in this work, we will disclose an additional principle direction in the SLBP, which will be referred to the third principle (TP) direction.  The Young's modulus has the maximum value in the TP direction, while the armchair and zigzag directions have the minimum Young's modulus.

In this paper, using the valence force field model (VFFM), we derive an analytic formula for the directional dependence of the Young's modulus in SLBP. Besides armchair and zigzag directions, we reveal the TP direction, in which a maximum Young's modulus value is reached.

%\section{Structure and potential}

\begin{figure}[tb]
  \begin{center}
    \scalebox{1}[1]{\includegraphics[width=8cm]{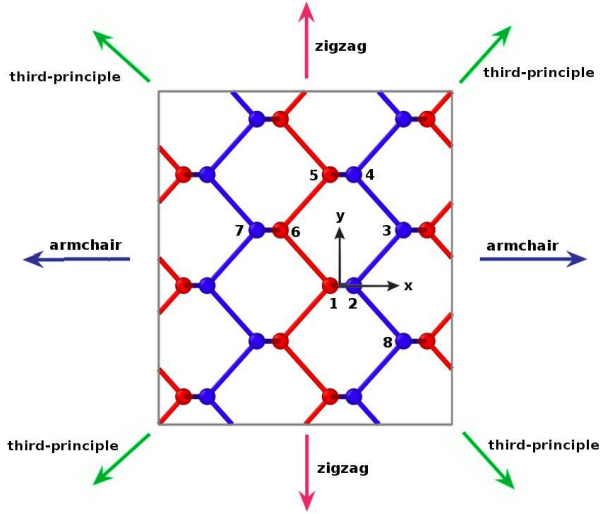}}
  \end{center}
  \caption{(Color online) SLBP structure. There are three principle directions, i.e., armchair (blue arrows), zigzag (red arrows), and the TP (green arrows) directions. Color is with respective to the atomic z-coordinate.}
  \label{fig_cfg}
\end{figure}

The atomic configuration of the SLBP is shown in Fig.~\ref{fig_cfg}. The structure parameters were measured in the experiment.\cite{TakaoY1981physica} Two in-plane lattice constants are $a_1=r_{37}=4.376$~{\AA} and $a_2=r_{24}=3.314$~{\AA}. The out-of-plane lattice constant is $a_3=10.478$~{\AA}. The origin of the Cartesian coordinate system is in the middle of $\vec{r}_{12}$. The x-axis is in the horizontal direction and the y-axis is in the vertical direction. There are four inequivalent atoms in the unit cell $\vec{a}_1\times\vec{a}_2$ of the SLBP, which will be chosen as atoms 1, 2, 3, and 6 in this work. The coordinate of these atoms are $\vec{r}_1=(-ua_1, 0, -va_3)$, $\vec{r}_2=(ua_1, 0, va_3)$, $\vec{r}_3=(0.5a_1-ua_1, 0.5a_2, va_3)$, and $\vec{r}_6=(-0.5a_1+ua_1, 0.5a_2, -va_3)$. The two dimensionless parameters are $u=0.0806$ and $v=0.1017$. The bond lengths from the experiment are $d_1=r_{23}=r_{16}=2.2449$~{\AA} and $d_2=r_{12}=2.2340$~{\AA}, and the two angles are $\theta_1=\theta_{328}=0.535\pi$ and $\theta_2=\theta_{321}=0.567\pi$.

\begin{table*}
\caption{ Parameters (in eV\AA$^{-2}$ ) for the VFFM potential from Ref~\onlinecite{KanetaC1982ssc}.}
\label{tab_vffm}
\begin{tabular*}{\textwidth}{@{\extracolsep{\fill}}|c|c|c|c|c|c|c|c|c|}
\hline 
$K_{r}$ & $K_{r}'$ & $K_{\theta}$ & $K_{\theta}'$ & $K_{rr'}$ & $K_{rr'}'$ & $K_{r\theta}$ & $K_{r\theta}'$ & $K''_{r\theta}$\tabularnewline
\hline 
9.9715 & 9.4598 & 1.0764 & 0.9341 & 1.1057 & 1.1057 & 0.7207 & 0.7207 & 0.7207\tabularnewline
\hline 
\end{tabular*}
\end{table*}

Several empirical potentials have been developed to describe the atomic interaction for the SLBP, including the VFFM potential\cite{KanetaC1982ssc} and the Stillinger-Weber potential.\cite{JiangJW2013sw} Both potentials were fitted to the phonon dispersion of the SLBP. The Stillinger-Weber potential includes some nonlinear properties, so it can be applied in molecular dynamics simulations of the SLBP. The VFFM is a linear model, so it is suitable for the investigation of linear properties in the SLBP, like the elastic bending modulus studied in this work. The VFFM is convenient for deriving analytic expressions for elastic properties thanks to its simplicity. An analytic expression is of help for an explicit understanding of the elastic properties. Hence, we will apply the VFFM to derive an analytic formula for the Young's modulus of the SLBP.

There are nine terms in the VFFM potential,
\begin{eqnarray}
V_{r} & = & \frac{1}{2}K_{r}\left(\Delta d_{1}\right)^{2};\\
\label{eq_Vr}
V_{r}' & = & \frac{1}{2}K_{r}'\left(\Delta d_{2}\right)^{2};\\
V_{\theta} & = & \frac{1}{2}K_{\theta}d_{1}^{2}\left(\Delta\theta_{1}\right)^{2};\\
V_{\theta}' & = & \frac{1}{2}K_{\theta}'d_{1}d_{2}\left(\Delta\theta_{2}\right)^{2};\\
V_{rr'} & = & \frac{1}{2}K_{rr'}\left(\Delta d_{1}\right)\left(\Delta d_{1}\right);\\
V_{rr'}' & = & \frac{1}{2}K_{rr'}'\left(\Delta d_{1}\right)\left(\Delta d_{2}\right);\\
V_{r\theta} & = & \frac{1}{2}K_{r\theta}d_{1}\left(\Delta d_{1}\right)\left(\Delta\theta_{1}\right);\\
V_{r\theta}' & = & \frac{1}{2}K_{r\theta}'\sqrt{d_{1}d_{2}}\left(\Delta d_{1}\right)\left(\Delta\theta_{2}\right);\\
V''_{r\theta} & = & \frac{1}{2}K''_{r\theta}\sqrt{d_{1}d_{2}}\left(\Delta d_{2}\right)\left(\Delta\theta_{2}\right).
\label{eq_Vrtpp}
\end{eqnarray}
The VFFM describes the energy variation of the system due to a small change in the bond length ($\Delta b_i$) and the angle ($\Delta \theta_i$) with $i=1,2$, which are induced by strain in the present work. The $V_r$ term describes the bond stretching energy for intra-group bond lengths like $r_{23}$. The $V_r'$ term is the energy corresponding to the bond stretching for inter-group bond lengths like $r_{12}$. The $V_{\theta}$ term describes the energy associating with the variation of intra-group angles like $\theta_{234}$. The $V'_{\theta}$ term describes the energy variation due to the variation of the inter-group angles like $\theta_{123}$. The $V_{rr'}$ term describes the potential energy for the simultaneous variation of two different intra-group bonds like $r_{23}$ and $r_{24}$. The $V'_{rr'}$ term gives the potential energy for the simultaneous variation of bonds like $r_{21}$ and $r_{23}$. The $V_{r\theta}$ term is for the energy association with the simultaneous variation of an intra-group bond like $r_{32}$ and an intra-group angle like $\theta_{234}$. The $V'_{r\theta}$ term gives the potential energy for the simultaneous variation of an inter-group angle like $\theta_{123}$ and an intra-group bond like $r_{23}$.  The $V''_{r\theta}$ term gives the potential energy for the simultaneous variation of an inter-group angle like $\theta_{123}$ and an inter-group bond like $r_{12}$. All parameters are shown in Tab.~\ref{tab_vffm}. The unit of these parameters has been converted from dyne/cm in the original work to eV\AA$^{-2}$.

To compute the Young's modulus, the uniaxial uniform strain $\vec{\epsilon}$ is applied to the structure. The magnitude of the strain is $\epsilon$ and the direction of the strain is $\hat{\epsilon}=(\cos\phi, \sin\phi, 0)$. The angle $\phi$ is counted from the x-axis. Based on the VFFM potential, the strain energy density is

\begin{eqnarray}
 W\times S_{0} & = & 2\frac{K_{r}}{2}\left(\Delta r_{23}\right)^{2}+2\frac{K_{r}}{2}\left(\Delta r_{28}\right)^{2}+2\frac{K_{r}'}{2}\left(\Delta r_{21}\right)^{2}\nonumber\\
 & + & 4\frac{K_{\theta}}{2}\left(\Delta\theta_{328}\right)^{2}+4\frac{K_{\theta}'}{2}\left(\Delta\theta_{321}\right)^{2}\nonumber\\
 & + & 4\frac{K_{\theta}'}{2}\left(\Delta\theta_{821}\right)^{2}+4\frac{K_{rr'}}{2}\left(\Delta r_{23}\Delta r_{28}\right)\nonumber\\
 & + & 4\frac{K_{rr'}'}{2}\left(\Delta r_{23}\Delta r_{21}\right)+4\frac{K_{rr'}'}{2}\left(\Delta r_{28}\Delta r_{21}\right)\nonumber\\
 & + & 4\frac{K_{r\theta}}{2}\left(\Delta r_{23}\Delta\theta_{234}\right)+4\frac{K_{r\theta}}{2}\left(\Delta r_{43}\Delta\theta_{234}\right)\nonumber\\
 & + & 4\frac{K_{r\theta}'}{2}\left(\Delta r_{23}\Delta\theta_{123}\right)+4\frac{K_{r\theta}'}{2}\left(\Delta r_{28}\Delta\theta_{128}\right)\nonumber\\
 & + & 4\frac{K_{r\theta}''}{2}\left(\Delta r_{12}\Delta\theta_{123}\right)+4\frac{K_{r\theta}''}{2}\left(\Delta r_{12}\Delta\theta_{128}\right),
\end{eqnarray}
where $S_{0}=a_{1}\times a_{2}$ is the area of the unit cell $\vec{a}_{1}\times\vec{a}_{2}$. The right-hand side gives the total VFFM energy for a unit cell.

%\section{Young's modulus}

The Young's modulus can be obtained through its definition,
\begin{eqnarray}
E & =& \frac{\partial^{2}W}{\partial\epsilon^{2}}\nonumber\\
 & = & \frac{1}{S_{0}}[2K_{r}\left(\alpha_{1}^{2}+\alpha_{3}^{2}\right)+2K_{r}'\alpha_{2}^{2}\nonumber\\
 & + & 4K_{\theta}\beta_{1}^{2}+4K_{\theta}'\left(\beta_{2}^{2}+\beta_{4}^{2}\right)+4K_{rr'}\alpha_{1}\alpha_{3}\nonumber\\
 & + & 4K_{rr'}'\alpha_{2}\left(\alpha_{1}+\alpha_{3}\right)+4K_{r\theta}\left(\alpha_{1}+\alpha_{3}\right)\beta_{1}\nonumber\\
 & + & 4K_{r\theta}'\left(\alpha_{1}\beta_{2}+\alpha_{3}\beta_{4}\right)+4K_{r\theta}''\alpha_{2}\left(\beta_{2}+\beta_{4}\right)],
\label{eq_E}
\end{eqnarray}
where the strain-induced variations for the bond length and the angle have been expressed as linear functions of strain; i.e., $\Delta d_i=\alpha_i\epsilon$ and $\Delta \theta_i=\beta_i\epsilon$, with $i=1, 2, 3, 4$. We have introduced six geometrical coefficients,
\begin{eqnarray}
\alpha_{1} & = & \frac{\partial r_{23}}{\partial\epsilon}|_{\epsilon=0};\alpha_{2}=\frac{\partial r_{12}}{\partial\epsilon}|_{\epsilon=0};\alpha_{3}=\frac{\partial r_{28}}{\partial\epsilon}|_{\epsilon=0};\\
\beta_{1} & = & \frac{\partial\theta_{328}}{\partial\epsilon}|_{\epsilon=0};\beta_{2}=\frac{\partial\theta_{321}}{\partial\epsilon}|_{\epsilon=0};\beta_{4}=\frac{\partial\theta_{821}}{\partial\epsilon}|_{\epsilon=0}.
\end{eqnarray}

The x-axis is rotated to the strain direction $\hat{\epsilon}$. The coordinate for the vector, $\vec{r}=(x,y,z)$, in this new coordinate system is
\begin{eqnarray}
\left(\begin{array}{c}
x_{\phi}\\
y_{\phi}\\
z_{\phi}
\end{array}\right) & = & \left(\begin{array}{ccc}
\cos\phi & \sin\phi & 0\\
-\sin\phi & \cos\phi & 0\\
0 & 0 & 1
\end{array}\right)\left(\begin{array}{c}
x\\
y\\
z
\end{array}\right),
\end{eqnarray}
where the under script $\phi$ denotes the new coordinate system with x-axis along the strain direction. In the new coordinate system, the strain is applied along the x-direction, so the effect of the strain is as follows
\begin{eqnarray}
\left(\begin{array}{c}
x_{\epsilon}\\
y_{\epsilon}\\
z_{\epsilon}
\end{array}\right) & = & \left(\begin{array}{ccc}
1+\epsilon & 0 & 0\\
0 & 1 & 0\\
0 & 0 & 1
\end{array}\right)\left(\begin{array}{c}
x_{\phi}\\
y_{\phi}\\
z_{\phi}
\end{array}\right),
\end{eqnarray}
where the under script $\epsilon$ denotes the coordinate under strain. The first derivative of the vector is $\frac{\partial\vec{r}_{\epsilon}}{\partial\epsilon}=x_{\phi}\hat{e}_{x\phi}$. As a result, the first derivative of the bond length is $\frac{\partial r_{\epsilon}}{\partial\epsilon}|_{\epsilon=0}=\frac{1}{r_\epsilon}\vec{r}_\epsilon\cdot\frac{\partial\vec{r}_\epsilon}{\partial\epsilon}=\frac{x_{\phi}^{2}}{r}$.

For the bond length $d_{1}=r_{23}$, we have $x_{23\phi}=\left(0.5-2u\right)a_{1}\cos\phi+0.5a_{2}\sin\phi$, so we get
\begin{eqnarray}
\alpha_{1} & = & \frac{\partial r_{23}}{\partial\epsilon}|_{\epsilon=0}=\frac{1}{r_{23}}x{}_{23\phi}^{2}\nonumber\\
 & = & \frac{1}{d_{1}}\left[\left(0.5-2u\right)a_{1}\cos\phi+0.5a_{2}\sin\phi\right]^{2}.
\end{eqnarray}
For the bond length $d_{2}=r_{21}$, we have $x_{21\phi}=-2ua_{1}\cos\phi$, so we get 
\begin{eqnarray}
\alpha_{2} & = & \frac{\partial r_{21}}{\partial\epsilon}=\frac{1}{r_{21}}x{}_{21\phi}^{2}=\frac{1}{d_{2}}\left(2ua_{1}\cos\phi\right)^{2}.
\end{eqnarray}
After the strain is applied, it is obvious that $r_{23}\not=r_{28}$, so we have another bond length in the deformed SLBP, i.e., $d_{3}=r_{28}$. We have $x_{28\phi}=\left(0.5-2u\right)a_{1}\cos\phi-0.5a_{2}\sin\phi$, which leads to
\begin{eqnarray}
\alpha_{3} & = & \frac{\partial r_{28}}{\partial\epsilon}|_{\epsilon=0}=\frac{1}{r_{28}}x{}_{28\phi}^{2}\nonumber\\
 & = & \frac{1}{d_{1}}\left[\left(0.5-2u\right)a_{1}\cos\phi-0.5a_{2}\sin\phi\right]^{2}.
\end{eqnarray}

\begin{figure}[tb]
  \begin{center}
    \scalebox{1.0}[1.0]{\includegraphics[width=8cm]{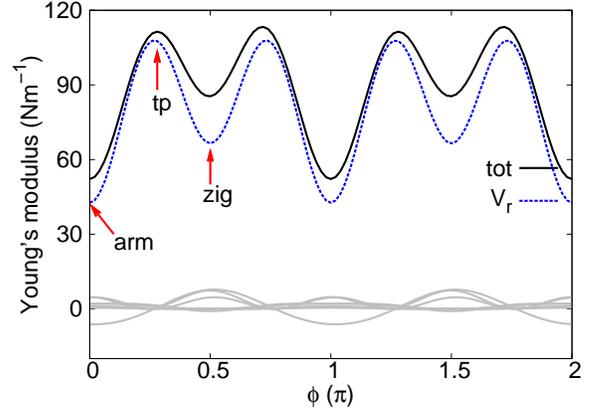}}
  \end{center}
  \caption{(Color online) Direction-dependent Young's modulus for SLBP. The total Young's modulus (black solid line) is mainly contributed by the $V_r$ potential term (blue doted line). The other eight potential terms (gray lines) contribute less than 10\% to the total Young's modulus. The maximum Young's modulus is in the TP direction at $\phi_{\rm tp}=0.268\pi$.}
  \label{fig_young}
\end{figure}

We consider the strain induced variation for the angle $\theta_{1}=\theta_{328}$. According to the definition, $\cos\theta_{328}=\hat{n}_{23}\cdot\hat{n}_{28}$, we have
\begin{eqnarray}
\beta_{1} & = & \frac{\partial\theta_{1}}{\partial\epsilon}=-\frac{1}{\sin\theta_{1}}\frac{\partial}{\partial\epsilon}\cos\theta_{1}\nonumber\\
 & = & -\frac{1}{\sin\theta_{1}}\left(\frac{2x_{23\phi}x_{28\phi}}{d_{1}^{2}}-\frac{2\alpha_{1}}{d_{1}}\cos\theta_{1}\right).
\end{eqnarray}
Analogous derivation gives
\begin{eqnarray}
\beta_{2} & = & \frac{\partial\theta_{2}}{\partial\epsilon}=\frac{\partial\theta_{321}}{\partial\epsilon}\nonumber\\
 & = & -\frac{1}{\sin\theta_{2}}\left[\frac{2x_{21\phi}x_{23\phi}}{d_{1}d_{2}}-\left(\frac{\alpha_{1}}{d_{1}}+\frac{\alpha_{2}}{d_{2}}\right)\cos\theta_{2}\right].
\end{eqnarray}
We find that $\theta_{321}\not=\theta_{821}$ in the deformed SLBP. As a result, we have
\begin{eqnarray}
\beta_{4} & = & \frac{\partial\theta_{4}}{\partial\epsilon}=\frac{\partial\theta_{821}}{\partial\epsilon}\nonumber\\
 & = & -\frac{1}{\sin\theta_{2}}\left[\frac{2x_{21\phi}x_{28\phi}}{d_{1}d_{2}}-\left(\frac{\alpha_{3}}{d_{1}}+\frac{\alpha_{2}}{d_{2}}\right)\cos\theta_{2}\right].
\end{eqnarray}

Inserting the above geometrical coefficients into Eq.~(\ref{eq_E}), we obtain the Young's modulus for the SLBP. Fig.~\ref{fig_young} shows the directional dependence for the Young's modulus. The contribution to the Young's modulus from all of the nine VFFM potential terms are displayed by individual curves. The Young's modulus in the armchair direction is is much less than the Young's modulus in the zigzag direction. Similar anisotropy in the Young's modulus has also been reported in several previous studies,\cite{AppalakondaiahS2012prb,QiaoJ2014nc,JiangJW2014bpnpr,JiangJW2014bpyoung,QinGarxiv14060261} though the obtained values show variability between the different studies. The difference is probably due to different computational methods and potentials that have been used in different studies. In present work, the Young's modulus is 52.2~{Nm$^{-1}$} in the armchair direction and 85.4~{Nm$^{-1}$} in the zigzag direction, which is within the range of the previously reported values.\cite{AppalakondaiahS2012prb,QiaoJ2014nc,JiangJW2014bpyoung,QinGarxiv14060261,WeiQ2014apl}

Fig.~\ref{fig_young} shows that the Young's modulus has a minimum value at $\phi=0$ (armchair direction). However, this curve clearly demonstrates that the Young's modulus actually has a minimum point at $\phi=\pi/2$ (zigzag direction), although the zigzag direction has larger Young's modulus value than the armchair direction. The maximum Young's modulus (111.4~{Nm$^{-1}$}) is in the direction with $\phi_{\rm tp}=0.268\pi$. We call this additional principle direction as the TP direction. 

\begin{figure}[tb]
  \begin{center}
    \scalebox{1.0}[1.0]{\includegraphics[width=8cm]{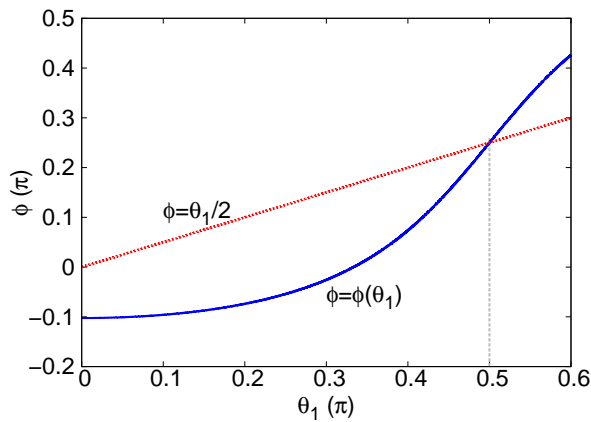}}
  \end{center}
  \caption{(Color online) Illustration for $\phi_{\rm tp}\not=\frac{\theta_1}{2}$. The cross-over between $\phi=\phi(\theta_1)$ (blue solid line) from Eq.~(\ref{eq_tp}) and $\phi=\frac{\theta_1}{2}$ (red dashed line) happens at $\theta_1=0.5\pi$. This is the only point satisfying $\phi_{\rm tp}=\frac{\theta_1}{2}$.}
  \label{fig_tp}
\end{figure}

The Young's modulus is mainly contributed by the $V_r$ term. We thus examined the Young's modulus computed from the $V_r$ term as
\begin{eqnarray}
E_{r} & = & \frac{2K_{r}}{S_{0}}\left(\alpha_{1}^{2}+\alpha_{3}^{2}\right)\nonumber\\
 & = & \frac{4K_{r}}{S_{0}d_{1}^{2}}\left[\left(0.5-2u\right)a_{1}\right]^{4}\left(\frac{1}{1+\tan^{2}\phi}\right)^{2}\nonumber\\
 &  & \left(1+6\tan^{2}\frac{\theta_{1}}{2}\tan^{2}\phi+\tan^{4}\frac{\theta_{1}}{2}\tan^{4}\phi\right).
\end{eqnarray}
As a result, the extreme points are determined by the condition,~{$\frac{\partial E_{r}}{\partial x}=0$}, which leads to
\begin{eqnarray}
\label{eq_arm}
\tan\phi & = & 0;\\
\label{eq_zig}
\tan\phi & = & \infty;\\
\tan^{2}\phi & = & \frac{1-3\tan^{2}\frac{\theta_{1}}{2}}{\tan^{2}\frac{\theta_{1}}{2}\left(\tan^{2}\frac{\theta_{1}}{2}-3\right)}.
\label{eq_tp}
\end{eqnarray}
These equations determine three principle directions in the SLBP. Eqs.~(\ref{eq_arm}) and ~(\ref{eq_zig}) give two local minimum for the Young's modulus in the direction with $\phi=0$ (armchair) or $\phi=\pi/2$ (zigzag). Eq.~(\ref{eq_tp}) determines the TP direction with $\phi_{\rm tp}=\pm0.268\pi$, in which the Young's modulus has the maximum value.

We note that the TP direction is almost coincident with the bond direction like $\vec{r}_{23}$ in Fig.~\ref{fig_cfg}, i.e., $\phi_{\rm tp}\approx \frac{\theta_1}{2}$. However, there is no guarantee for this equality according to Eq.~(\ref{eq_tp}). This can be more explicitly illustrated in Fig.~\ref{fig_tp}, which shows the functions $\phi=\phi(\theta_1)$ (blue solid line) from Eq.~(\ref{eq_tp}) and $\phi=\frac{\theta_1}{2}$ (red dashed line). The crossover between these two curves yields $\theta_1=0.5\pi$, which is the only $\theta_1$ value satisfying $\phi_{\rm tp} = \frac{\theta_1}{2}$. We actually have $\theta_1=0.535\pi$ for SLBP, which is not equal to $0.5\pi$, but these two values are very close to each other. As a result, the TP direction is very close to the bond direction like $\vec{r}_{23}$.

%\section{conclusion}
In conclusion, we have derived an analytic expression for directional-dependent Young's modulus of SLBP. The Young's modulus has the minimum value in the armchair direction, but the maximum Young's modulus is not in the zigzag direction. Instead, we find the TP direction in the SLBP with $\phi=0.268\pi$, along which the Young's modulus is maximum.

\textbf{Acknowledgements} The work is supported by the Recruitment Program of Global Youth Experts of China and the start-up funding from Shanghai University.

%\bibliographystyle{aipnum4-1}
%\bibliography{/home/JiangJinWu/Documents/papers/mypapers/latex/biball}
\end{document}